\documentclass[letterpaper,twocolumn,10pt]{article}
\usepackage{usenix}

\usepackage[utf8]{inputenc}
\usepackage{totcount}
\usepackage{url}
\usepackage{libertine}
\usepackage{listings}
\usepackage[T1]{fontenc}
\usepackage{xcolor}
\usepackage{quoting}
\usepackage{multirow}
\usepackage{rotating}
\usepackage{enumitem}
\usepackage{algorithm}
\usepackage[noend]{algpseudocode}
\usepackage{times} 
\usepackage{makecell}
\usepackage{xspace}

\newtoggle{shownotes}
\toggletrue{shownotes}
\newtotcounter{numnotes}

\newtoggle{showtodos}
\toggletrue{showtodos}
\newtotcounter{numtodos}

\newcommand{\system}{Cloudflow}
\newcommand{\cloudburst}{Cloudburst}
\newcommand{\ftable}{\texttt{Table}}
\newcommand{\fdataflow}{\texttt{Dataflow}}
\newcommand{\foperator}{\texttt{Operator}}
\newcommand{\map}{\texttt{map}}
\newcommand{\filter}{\texttt{filter}}
\newcommand{\groupby}{\texttt{groupby}}
\newcommand{\agg}{\texttt{agg}}
\newcommand{\join}{\texttt{join}}
\newcommand{\union}{\texttt{union}}
\newcommand{\lookup}{\texttt{lookup}}
\newcommand{\fuse}{\texttt{fuse}}
\newcommand{\anyof}{\texttt{anyof}}

\newcommand{\figwidth}{.48\textwidth}

\newcommand{\checkforcrud}{
  \ifboolexpr{test{\ifnumcomp{\totvalue{numnotes}}{>}{0}} or
              test{\ifnumcomp{\totvalue{numtodos}}{>}{0}}}{
    \pagecolor{red!20}
  }{}
}

\newcommand{\smallitem}[1]{\vspace{0.3em}\noindent\textbf{#1}}

\definecolor{deepblue}{rgb}{0,0,0.5}
\definecolor{deepred}{rgb}{0.6,0,0}
\definecolor{deepgreen}{rgb}{0,0.5,0} 
\definecolor{background}{rgb}{0.94,0.95,0.96}

\newcommand\pythonstyle{\lstset{
language=Python,
otherkeywords={self, def, CloudburstClient, CloudburstReference},             
keywordstyle=\color{deepblue},
backgroundcolor=\color{background},
emph={CloudburstClient},          
emphstyle=\color{deepred},    
stringstyle=\color{deepgreen},
showstringspaces=false,            %
basicstyle=\linespread{1.0}\ttfamily\scriptsize,
breaklines=true,
numbers=left,
numberstyle=\ttfamily,
columns=fixed,
basewidth=0.52em,
xleftmargin=6mm,
xrightmargin=0mm,
numberblanklines=true,
}}

\let\origthelstnumber\thelstnumber
\makeatletter
\newcommand*\Suppressnumber{%
  \lst@AddToHook{OnNewLine}{%
    \let\thelstnumber\relax%
     \advance\c@lstnumber-\@ne\relax%
    }%
}

\newcommand*\Reactivatenumber[1]{%
  \setcounter{lstnumber}{\numexpr#1-1\relax}
  \lst@AddToHook{OnNewLine}{%
   \let\thelstnumber\origthelstnumber%
   \refstepcounter{lstnumber}
  }%
}

\lstnewenvironment{python}[1][]
{
\pythonstyle
\lstset{#1}
}
{}

\begin{document}

\date{}

\title{\Large \bf Optimizing Prediction Serving on Low-Latency Serverless Dataflow}

\author{
{\rm Vikram Sreekanti} \\ UC Berkeley
\and
{\rm Harikaran Subbaraj} \\ UC Berkeley
\and
{\rm Chenggang Wu} \\ UC Berkeley
\and
{\rm Joseph E. Gonzalez} \\ UC Berkeley
\and
{\rm Joseph M. Hellerstein} \\ UC Berkeley
}

\maketitle

\checkforcrud{}

\begin{abstract}
    \emph{Prediction serving} systems are designed to provide large volumes of
    low-latency inferences from machine learning models. 
    These systems mix data processing and computationally intensive model inference,
    and benefit from multiple heterogeneous processors and distributed computing resources.
    In this paper, we argue that a familiar dataflow API is well-suited to this latency-sensitive task, 
    and amenable to optimization even with unmodified black-box ML models. We 
    present the design of \emph{\system{}}, a system that provides such an API 
    and realizes it 
    on an autoscaling serverless back-end. \system{} transparently implements performance-critical optimizations including operator fusion and competitive execution.
    %
    Our evaluation shows that \system{}'s optimizations yield significant performance improvements on synthetic workloads and that \system{} outperforms state-of-the-art prediction serving systems by as much as 2$\times$ on real-world prediction pipelines, meeting latency goals of demanding applications like real-time video analysis.
\end{abstract}
\section{Introduction} \label{sec:intro}

Machine learning has become ubiquitous over the last decade in an increasingly broad set of fields, ranging from manufacturing to medicine and sports.
Much of the systems research surrounding machine learning has focused on improving the infrastructure that supports the creation of models---tools like Tensorflow~\cite{abadi2016tensorflow}, PyTorch~\cite{pytorch}, and MLLib~\cite{meng2016mllib} have greatly simplified the process of developing and training models at scale.
The models trained in these systems are deployed in numerous applications---generating social media feeds, enabling chat bots, designing video games, and so on.


\begin{figure} 
  \centering
    \includegraphics[height=1.2in]{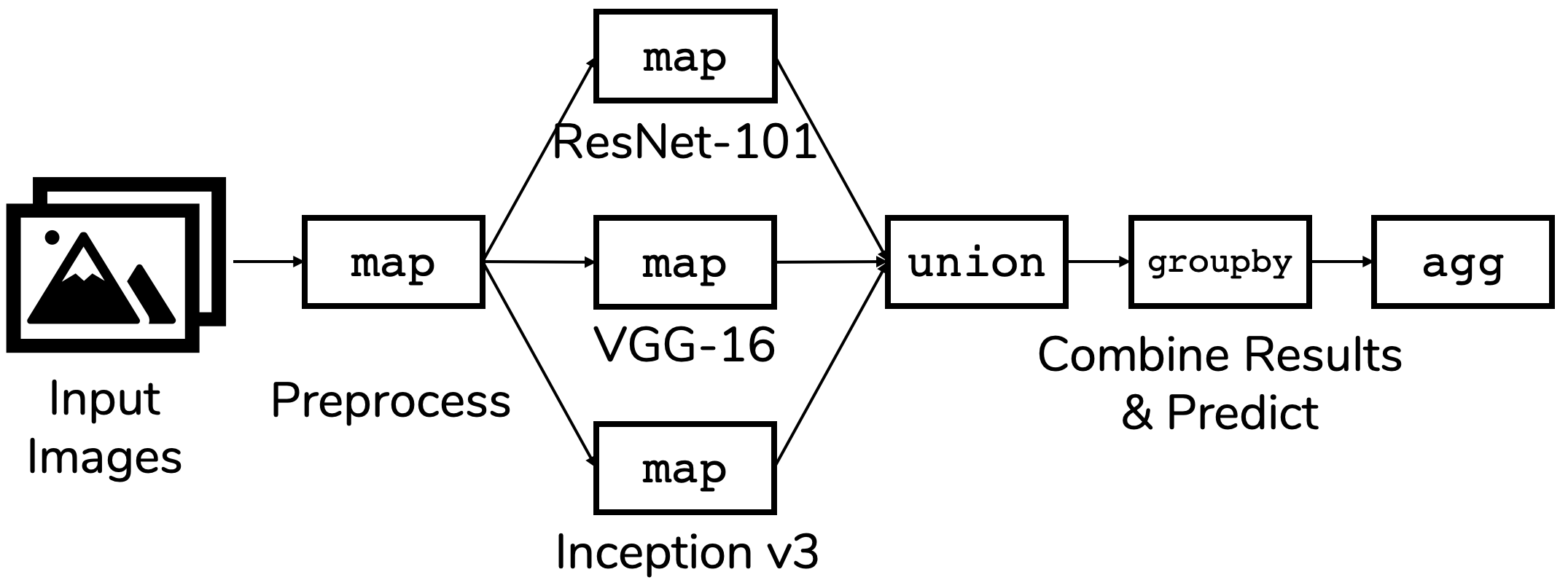}
      \begin{python}
 fl = cloudflow.Dataflow([('url', str)])
 img = fl.map(img_preproc)
 p1 = img.map(resnet_101)
 p2 = img.map(vgg_16)
 p3 = img.map(inception_v3)
 fl.output = p1.union(p2,p3).groupby(rowID).agg(max,'conf')
    \end{python}
  \caption{\small
    An example prediction serving pipeline to classify a set of images using an ensemble of three models, and the \system{} code to specify it.
    The models are run in parallel; when all finish, the result with the highest confidence is output.
  }
  \label{fig:example}
  \label{lst:example}
\end{figure}



This process of deploying a trained model for use in a larger application is often called \emph{prediction serving}.
Prediction serving is a particularly interesting task because it combines the complex computations of machine learning with the performance requirements of interactive applications.
More specifically, it has three key properties: (1) it is \emph{computationally intensive}; (2) it is a part of an interactive application, meaning it has \emph{low latency requirements}; and (3) it is \emph{compositional}, meaning a single request passes through multiple stages.
Figure~\ref{fig:example} shows an example prediction serving pipeline, which normalizes an input image, runs three image classification models in parallel, and combines the results to make a prediction (often called a \emph{model ensemble}).

Commercial services like AWS Sagemaker and Azure ML have emerged in recent years to attempt to fill this gap.
These systems deploy individual models as separate microservices which enables neat modularity of pipeline stages.
However, this approach gives the system no visibility into the \emph{structure} of the computation and how the individual microservices relate to each other, which significantly complicates debugging and limits end-to-end performance optimization.


Alternately, Pretzel~\cite{lee2018pretzel} is a recent research system which explored fine-grained optimizations of \emph{white-box}, dataflow-style prediction pipelines that leverage full visibility into the pipeline and the semantics of each stage.
However, this approach also requires the developer to rewrite their \emph{individual models} (i.e., pipeline stages) in Pretzel's DSL, which is cumbersome and limits model development.
It also makes Pretzel inflexible to innovations in machine learning frameworks.

In this paper, we argue for a natural middle ground: a simple dataflow API for prediction pipelines based on arbitrary (black-box) operators,
which are typically models trained by users in their library of choice (e.g., TensorFlow, PyTorch, Scikit-Learn).
Rather than requiring models to be rewritten in a white-box fashion, a graph of familiar dataflow operators (e.g., \map{}, \filter{}, \join{}) can be used to wrap black-box models. 
The dataflow API also provides visibility into the structure of the pipeline, which allows for the end-to-end computation to be optimized---e.g., two operators that have large data dependencies might be fused together.

Existing large-scale distributed dataflow systems are not well suited for prediction serving for two reasons.
First, prediction serving tasks have low latency requirements.
Systems like Apache Spark~\cite{zaharia2012resilient} and Apache Flink~\cite{carbone2015apache}, however, are optimized for throughput rather than for minimizing tail-latency.

Second, prediction serving systems---like all interactive applications---must operate in the presence of bursty and unpredictable workloads~\cite{facebook,shahrad2020serverless}, which requires fine-grained resource allocation.
Unfortunately, efforts~\cite{ds2, autoscale, nephele, fu:drs2017, lohrmann15, kalavri18} 
to address operator-level auto-scaling have primarily focused on meeting throughput requirements. 

We present \system{}, a dataflow system for prediction serving pipelines.
\system{} is built on top of \cloudburst{}~\cite{sreekanti2020cloudburst}, a stateful serverless programming platform.
\cloudburst{} is a Functions-as-a-Service (FaaS) system that provides low-latency access to state by enabling function composition, message passing, and caching of frequently accessed data.
Layering on top of a stateful serverless platform enables fine-grained resource allocation, provides a simple functional programming model to match dataflow, and supports efficient state retrieval.

\system{} exposes a dataflow API that enables simple composition patterns common in prediction serving.
The dataflow model enables us to apply both dataflow and prediction serving optimizations (e.g., operator fusion, competitive execution) to optimize those pipelines, even while treating models as black boxes.
We demonstrate that these optimizations can be applied without user intervention.
In practice, \system{} is able to outperform prior systems for prediction serving by as much as 2$\times$, and—--critically—--meets latency goals of demanding applications like real-time video analysis.

In sum, the contributions of this paper are as follows:

\begin{itemize}

    \item The modeling and implementation of prediction serving pipelines as dataflow graphs using a familiar API of functional operators. 
    This includes the implementation of common patterns like ensembles and cascades.
    
    \item Leveraging and extending a stateful serverless platform to enable efficient dataflows without changes to user programs, as discussed in Sections~\ref{sec:api} and~\ref{sec:opts}.
    
    \item The automatic application of well-studied optimizations from the dataflow and prediction serving domains to \system{} dataflows.
    
    \item A thorough evaluation of \system{}, demonstrating the benefits of optimized dataflow on both synthetic benchmarks and on real-world prediction pipelines.
    
\end{itemize}

\section{Background and Motivation} \label{sec:bg}

In this section, we discuss challenges posed by prediction serving (Section~\ref{sec:bg-predserv}).
We then argue that dataflow is well-suited to address these challenges and discuss how we optimize prediction pipelines (Section~\ref{sec:bg-opts}), and  we motivate the use of a serverless runtime and our choice, \cloudburst{} (Section~\ref{sec:bg-cloudburst}). 

\subsection{Prediction Serving} \label{sec:bg-predserv}

Prediction serving has become a common part of many tasks, such as organizing social media news feeds, responding to voice assistant queries (Alexa, Siri, etc.), detecting fraud in financial transactions, and analyzing radiological scans.
The models used in these applications, particularly with the advent of deep neural networks, can be incredibly computationally intensive~\cite{johnson2017google, hazelwood2018applied}.
As a result, it has become commonplace for these models to run on specialized hardware, like GPUs or Google's TPU (Tensor Processing Unit)~\cite{jouppi2017datacenter}.

Furthermore, these applications are often interactive and require predictions to be made within tight latency bounds. Facebook published a detailed case study~\cite{facebook} in which they describe models that query user-specific state and generate predictions in real-time---about a few hundred milliseconds.
Predictions that miss their latency deadlines are usually discarded in favor of a default response~\cite{zhou2018deep, he2012zeta}, making it critical to minimize both median \emph{and} tail (99th percentile) latencies. 

Importantly, it is typical for applications to compose multiple models to make a single prediction~\cite{lee2018pretzel}.
Consider the example in Figure~\ref{fig:example}.
This image classification pipeline first preprocesses and normalizes an image, and then runs three different models in parallel---each has different specializations. 
The results of all three models are aggregated to generate a prediction.
This particular pattern is called a \emph{model ensemble}.

\subsection{Dataflow and Optimizations} \label{sec:bg-opts}

The sequence of tasks in a prediction pipeline forms a natural dataflow graph: Each stage of the pipeline receives an input, applies a transformation, and passes the result downstream.
The output of the last stage is the prediction rendered by the whole pipeline.
Using a traditional dataflow model with operators like \map{}, \filter{}, and \join{} makes constructing pipelines easy---the ensemble in Figure~\ref{fig:example} being a case in point.
On inspection this may seem straightforward, but it is in stark contrast with existing systems: AWS Sagemaker forces users to manually construct containers for each model stage, while Pretzel~\cite{lee2018pretzel} requires users to rewrite their models in a custom API to leverage the system's optimizations.
In this paper, we argue that a traditional dataflow API enables us to implement a number of pipeline optimizations \emph{without} requiring any modification of black-box ML models.

%


We focus on five optimizations from the data processing and prediction serving literature that apply here~\cite{antoshenkov1993dynamic, bress2013time, crankshaw2017clipper, lee2018pretzel, taso, paritymodels}, but this list is not meant to be exhaustive.


\smallitem{Operator Fusion}. Separate stages in a dataflow may pass significant data like videos or images. To minimize communication, it can be beneficial to fuse these operators to avoid data movement.


\smallitem{Competitive Execution}. Machine learning models can have highly variable execution times~\cite{kosaian2019parity}, depending on model structure and input complexity.
Similar to straggler mitigation in MapReduce~\cite{dean2008mapreduce}, competitive execution of inference has been shown to improve tail latencies for ML models~\cite{sureshc3, paritymodels}.


\smallitem{Operator Autoscaling and Placement}. Given the diversity of compute requirements in a pipeline (e.g., simple relational operations vs neural-net inference), we often want to devote more---and more specialized---resources to bottleneck tasks.



\smallitem{Data Locality}. While prediction pipelines are typically computationally intensive tasks, they also often involve significant data access~\cite{gupta2020architectural}. 
Remote data access can easily become a bottleneck unless we optimize for 
data locality by placing computations near cached data.



\smallitem{Batching}. 
Large models like neural nets can benefit greatly from batched execution on 
vectorized processors like GPUs
~\cite{bianco2018benchmark, gao2018low}, resulting in better throughput at the cost of higher latency.
The improved throughput translates into better resource utilization and thereby reduces costs.

In Section~\ref{sec:opts} we explore each of these optimizations in more detail, and explain how we deliver on them in \system{}.

\subsection{Deploying Prediction Pipelines} \label{sec:bg-cloudburst}


Having motivated the use of dataflow to construct and optimize prediction pipelines, we now consider the runtime on which they are executed.
Two key concerns affect our decision: (1) tight latency constraints, and (2) nimble responses to unpredictable workload changes. 
Unfortunately, batch streaming dataflow systems (e.g., Apache Flink~\cite{carbone2015apache}, Apache Spark~\cite{zaharia2012resilient}) are unsuitable on both counts---they are throughput-oriented systems that do not scale easily.

Instead, we chose to use serverless Functions-as-a-Service (FaaS) infrastructure, primarily because FaaS systems natively support fine-grained elasticity---an integral part of operator autoscaling. 
Additionally, FaaS' functional programming style neatly matches the operator-based abstractions of dataflow---each operator can be deployed as a separate function.

Despite the benefits of fine-grained elasticity, commercial FaaS offerings have key shortcomings around data intensive workloads~\cite{hellerstein2018serverless}---they have high invocation latencies and force users into data-shipping anitpatterns.
To overcome these limitations, we turned to research platforms like Archipelago~\cite{singhvi2019archipelago}, OpenLambda~\cite{hendrickson2016serverless}, and Cloudburst~\cite{sreekanti2020cloudburst} as more performant alternatives. 
Of these, we chose Cloudburst as \system{}'s underlying compute engine because---in addition to standard FaaS autoscaling---it enables low-latency function composition and data access.
We provide a brief overview of the system here and refer the interested reader to~\cite{sreekanti2020cloudburst}. 

\cloudburst{} is a FaaS platform that supports stateful serverless programming with three kinds of state sharing: function composition, message passing, and data lookups in caches located on the same machines where code is run.
The system is built on top of Anna~\cite{wu2019anna, wu2019autoscaling}, a low-latency autoscaling key-value store.
\cloudburst{} layers a set of function executor nodes on top of Anna; each executor has multiple workers that respond to user requests as well as a cache that intermediates on KVS reads and writes.
The system optimizes for data locality by scheduling requests on executor nodes where input data is likely to be cached.
\cloudburst{} is able to outperform commercial FaaS platforms by orders of magnitude for stateful tasks and significantly cut data transfer costs.

In addition to the standard FaaS execution model, \cloudburst{} allows users to pre-register compositions of functions as directed acyclic graphs (DAGs); a whole DAG is scheduled and executed with a single user request.
Combined with the stateful architecture, this DAG-based programming model is a natural fit for deploying prediction dataflows.
While \cloudburst{} was a good starting point for our infrastructure, the system nonetheless had a number of key limitations that we had to overcome---we describe these in Section~\ref{sec:opts}.


\section{Architecture and API} \label{sec:api}

In this section, we describe the \system{} API for composing dataflows, how that API captures common prediction pipelines, and how \system{} pipelines are compiled down to \cloudburst{} serverless functions.


\subsection{Dataflow API} \label{sec:api-dataflow}

The \system{} API centers around three main concepts: a simple \underline{\ftable{}} type for data, computational \underline{\foperator{}s} that compute over \ftable{}s, and a functional-style \underline{\fdataflow{}} to author specifications of DAGs of \foperator{}s. 
A Python program can specify a \fdataflow{} and execute it on a \ftable{}; the \system{} runtime is responsible for taking that specification and invoking the \foperator{}s to generate an output \ftable{}.

\begin{table*}\footnotesize
\begin{center}
\begin{tabular}{|p{.1\textwidth}|p{.25\textwidth}|p{.20\textwidth}|p{.35\textwidth}|}
\hline

\textbf{API} & \textbf{Inputs} & \textbf{Output Type} & \textbf{Description} \\
\hline

\shortstack[l]{\map{}}
& \shortstack[l]{\texttt{fn: }$(c_1, ..., c_n) \rightarrow (d_1, ..., d_m)$, \\
    \texttt{table: Table}$[c_1, ..., c_n][column?]$}
& \shortstack[l]{\texttt{Table}$[d_1, ..., d_n][column?]$} 
& Apply a function \texttt{fn} to each row in \texttt{table} \\
\hline

\filter{} 
& \shortstack[l]{\texttt{fn: }$(c_1, ..., c_n) \rightarrow bool$, \\
    \texttt{table: Table}$[c_1, ..., c_n][column?]$} 
& \texttt{Table}$[c_1, ..., c_n][column?]$ 
& \shortstack[l]{Apply Boolean function to each row in \texttt{table} \\ and include only rows with \texttt{true} results} \\
\hline

\groupby{} 
& \shortstack[l]{\texttt{column: str}, \\
    \texttt{table: Table}$[c_1, ..., c_n][]$}
& \texttt{Table: }$[c_1, ..., c_n][column]$  
& \shortstack[l]{Group rows in an ungrouped \texttt{table} by the value \\ in \texttt{column}} \\
\hline

\agg{} 
& \shortstack[l]{\texttt{agg\_fn: }$\{c_i\} \rightarrow d$, \\
    \texttt{table: Table}$[c_1, ..., c_n][column?]$} 
& \texttt{\texttt{Table}$[column?, d][]$} 
& \shortstack[l]{Apply a predefined aggregate function \texttt{agg\_fn} \\ (count, sum, min, max, avg) to column $c_i$ of \texttt{table}} \\
\hline

\lookup{} 
& \shortstack[l]{\texttt{key: str?} $c_k$\texttt{?}, \\
    \texttt{Table}$[c_1, ..., c_n][column?]$}
& \texttt{Table}$[c_1, ..., c_n, key][column?]$ 
& \shortstack[l]{Retrieve an object from the underlying KVS and \\ insert into the \texttt{table}} \\
\hline

\join{} 
& \shortstack[l]{\texttt{left: Table}$[c_1, ..., c_n][]$, \\
    \texttt{right: Table}$[d_1, ..., d_m][]$, \\
    \texttt{key: str?} \\
    \texttt{how: left?, outer?}}
& \texttt{Table}$[c_1, ..., c_n, d_1, ... d_m][]$ 
& \shortstack[l]{Join two \ftable{}s on \texttt{key}, using the automatically \\ assigned query ID as a default. Optionally specify \\ left join or (full) outer join mode.}\\
\hline

\union{} 
& \texttt{\{Table$[c_1, ..., c_n][column?]$, ...\}}
& \texttt{Table}$[c_1, ..., c_n][column?]$ 
& \shortstack[l]{Form union of many \ftable{}s with matching schemas.}\\
\hline

\anyof{} 
& \texttt{\{Table$[c_1, ..., c_n][column?]$, ...\}}
& \texttt{Table}$[c_1, ..., c_n][column?]$ 
& \shortstack[l]{Pick any one of many \ftable{}s with matching schemas.}\\
\hline

\fuse{} 
& \shortstack[l]{\texttt{sub\_dag: Flow}, \\
    \texttt{table: Table}$[c_1, ..., c_n][column?]$} 
& \texttt{Table}$[d_1, ..., d_m][column?]$ 
& \shortstack[l]{An encapsulated chain of operators (see Section~\ref{sec:opts-fusion})}\\
\hline

\end{tabular}
\vspace{-2em}
\end{center}

\caption{\small
    The core \foperator{}s supported by \system{}.
    Each accepts a \ftable{} as input and and returns a \ftable{} as output. Our table type notation here is \texttt{Table}$[c_1, ..., c_n][column]$, where $c_1,\ldots,c_n$ is the schema, and $column$ is the grouping column. Optional items are labeled with a \texttt{?}.
}
\label{table:api}
\end{table*}

\begin{figure}[t]
\begin{python}
fl = cloudflow.Dataflow([('jpg_url', str)])
img = fl.map(img_preproc)
pred = img.map(resnet)
label = pred.map(convert_to_label) 
fl.output = label
fl.deploy()

input_tbl = \
  Table([('jpg_url', str)], \    
        ['s3://mybucket/cats.jpg', 's3://mybucket/dogs.jpg'])
out = fl.execute(input_tbl)
out.result()
\end{python}

\vspace{-1em}

\caption{A script to create a \system{} dataflow and execute it once.} 
\label{lst:example-code}
\end{figure}


The core data structure in \system{} is a simple in-memory relational \underline{\ftable{}}. 
A \ftable{} has a \emph{schema}, which is a list of column descriptors, each consisting of a name and an associated data type (e.g., \texttt{str}, \texttt{int}). It also has an optional \emph{grouping column}\footnote{
Our syntax naturally extends to support a list of grouping columns, but we omit that detail here for descriptive simplicity. Even without that syntax, one can form composite grouping columns with a \map{} operator.}.
We briefly defer our description of grouping columns to the discussion below of the dataflow operators in Table~\ref{table:api}.


A \system{} \underline{\fdataflow{}} represents a \emph{specification} for a DAG of dataflow operators with a distinguished input and output, as in Figure~\ref{fig:example}.
One can think of a \fdataflow{} as a declarative query, or a lazy functional expression, describing how to compute an output table from an input table. 
We step through Figure~\ref{lst:example-code} to illustrate.
A \fdataflow{} instance is instantiated with an input schema (Figure~\ref{lst:example-code}, line~1). To construct pipelines,
the \fdataflow{} class includes a set of method names that correspond one-to-one with the \foperator{} names in Table~\ref{table:api}; these methods generate new \fdataflow{} objects that append the specification of the corresponding operator onto their input dataflow. So, for example line~2 represents a dataflow that has a single-column table as its input, followed by the expression \texttt{map(img\_preproc)}; line~3 represents a dataflow equivalent to \texttt{flow.map(img\_preproc).map(resnet)}, and so on. \fdataflow{} \texttt{fl} is \emph{valid} after its \texttt{output} member is assigned (line~5). This assignment requires its right-hand-side to be derived from the same \fdataflow{} \texttt{fl}, representing a connected flow from the input to the output of \texttt{fl}.

To prepare a \fdataflow{} for execution, we use the \texttt{deploy} method (line 6), which compiles the \fdataflow{} and registers it with the \cloudburst{} runtime. 
Once deployed, we can repeatedly execute the \fdataflow{} by passing in a \ftable{} with the corresponding schema. 
\texttt{execute} immediately returns a Python future that represents the result table of the execution (line~11).
\cloudburst{} schedules and executes the DAG as described in~\cite{sreekanti2020cloudburst}, and results are stored in Anna.
The result is accessed by calling the \texttt{result} method of the future returned by \texttt{execute} (line~12).
During execution of a \fdataflow{}, each row in the input is automatically assigned a unique row ID, which stays with the row throughout execution.


\smallitem{Dataflow Operators}. The set of \system{} \foperator{}s is similar to the relational algebra, or the core of a dataframe 
API. It consists of a small set of dataflow operators, each of which takes in \ftable{}s and produces a \ftable{}. 
Table~\ref{table:api} provides an overview of the input/output types of each operator. 
The operators look familiar from batch processing and dataframe libraries, but recall that our focus here is on lightweight, interactive predictions over small in-memory request \ftable{}s.




The \map{}, \filter{}, \join{} and \union{} operators are fairly standard.
\groupby{} takes an ungrouped \ftable{} and returns a grouped \ftable{}.
\map, \filter, \union{}, \anyof{} and \fuse{} all accept \ftable{}s that can be grouped or ungrouped (denoted by $[column?]$) and return \ftable{}s with the same grouping.
\anyof{} passes exactly one of its input tables to the output; the decision is left up to the runtime~(Section~\ref{sec:opts-compete}).
\fuse{} is an internal operator that executes multiple other operators over a \ftable{}---we discuss its use in Section~\ref{sec:opts-fusion}.

The \agg{} operator supports basic aggregations: counts, sums, averages, min, and max. 
When \agg{} is applied over an ungrouped table, it returns a table with a single row; when applied over a grouped table, \agg{} returns an ungrouped table with one row per input group containing the group value and aggregate result.
\join{} takes two input tables, both of which must be ungrouped, and joins them on a user-specified key; if no key is specified, the two tables are joined on the row ID.
\system{} also supports left joins, where rows from the left table are included in the output even if there is no match in the other table, and outer joins, in which rows from each table are included in the output even if there are no matches. 

Finally, \lookup{} allows dataflows to interact with data outside the input table. Specifically it allows for reads from
the Anna KVS that \cloudburst{} uses for storage.
\lookup{}s can take either a constant or a column reference. If the input is a column reference $c_k$, the effect is akin to a join with the KVS: for each input row, \system{} retrieves the matching object in the KVS.
As we discuss in Section~\ref{sec:opts}, we use the information in the \lookup{} operator to take advantage of \cloudburst{}'s locality-aware scheduling.
In practice, many of the API calls in Table~\ref{table:api} have other arguments for user convenience (e.g., naming output columns) and to give \system{} hints about resource requirements and optimizations (e.g., requires a GPU, supports batching).
We omit describing these for brevity.

\smallitem{Typechecking and Constraints}. Similar to TFX~\cite{tfx}, \system{} supports basic typechecking to ensure the correctness of pipelines.
If the input type of an operator does not match the output type of the previous operator, \system{} raises an error.


To this end, we require programmers to provide Python type annotations for the functions they pass into \map{} and \filter{}.
Since Python is a weakly-typed language, these type annotations do not guarantee that the data returned by a function matches the annotation.
At runtime, the type of each function's output is inspected using Python's \texttt{type} operator.
Once again, if the type does not match the function's annotation, \system{} raises an error. 
This prevents Python from arbitrarily coercing types, causing pipelines to fail silently---generating incorrect or nonsensical results without surfacing an error.






\subsection{Prediction Serving Control Flow} \label{sec:api-control-flow}

Having described \system{}'s dataflow model, we show how it simplifies the implementation of control flow constructs common in prediction pipelines.
We describe some patterns and highlight their simplicity with code snippets.
This section is not meant to be exhaustive---rather we illustrate \system{}'s fitness for standard prediction serving control flow.

\smallitem{Ensembles}. In an ensemble pattern (e.g., Figure~\ref{fig:example}), an input is evaluated by multiple models in parallel.
After all models finish evaluating, the results are combined, either by picking the prediction with the highest confidence or by taking a weighted vote of the results.
Figure~\ref{lst:example} shows a \system{} implementation of a model ensemble.
After a preprocessing stage (line~2), three models are evaluated in parallel (lines~3-5), the results are unioned, and a final prediction is selected by using \texttt{agg} to pick the prediction with the maximum confidence (line~6).

\smallitem{Cascades}. In a cascade, models of increasing complexity are executed in sequence.
If an earlier model returns a prediction of sufficiently high confidence, then later models are skipped.
Figure~\ref{lst:cascade} shows the \system{} implementation of a model cascade.
After the simple model is executed (line~3), we retain rows with low confidence and apply the complex model to  them (line~4).
We then join the simple and complex models' results using a left join to include rows that were filtered out from the simple model (line~6). 
We then apply a \texttt{max\_conf} function to report the prediction with the highest confidence.

\begin{figure}[t]
\begin{python}
flow = cloudflow.Dataflow(['jpg_url', str]))
img = flow.map(preproc)
simple = img.map(simple_model)
complex = simple.filter(low_confidence).map(complex_model)

flow.output = simple.join(complex, how='left').map(max_conf)
\end{python}
\vspace{-1em}

\caption{A simple two-model cascade specified in \system{}.}
\label{lst:cascade}
\end{figure}

\subsection{Discussion}

In addition to the dataflow operators highlighted in Table~\ref{table:api}, \system{} also has an \texttt{extend} method.
\texttt{extend} takes in another valid flow as an argument and appends its DAG to the existing flow, creating a dataflow that chains the pair.
This allows for the easy composition of two flows. For example, multiple users in an organization might share an image preprocessing flow to which they each append a custom image classification flow.

In sum, \system{}'s API significantly simplifies the process of constructing prediction pipelines as compared to other systems.
Our dataflow model provides natural abstractions for \emph{composition}---e.g., chains of user-defined code in Figure~\ref{lst:example-code}, parallel executions with joins in Figures~\ref{lst:example} and Figure~\ref{lst:cascade}.
AWS Sagemaker has limited support for prediction pipelines---all operators must be in a sequence---and AzureML and Clipper have no support for multi-stage pipelines to our knowledge. 

Additionally, \system{} takes simple Python specs and automatically deploys pipelines for the user in a serverless framework. 
AzureML and Sagemaker require users to create custom containers for each stage\footnote{These systems will automatically generate a containerized version of the model only if you use their development infrastructure end-to-end.}.
As we discuss in Section~\ref{sec:eval-pipelines-setup}, porting pipelines to other systems required writing and deploying long-lived driver programs to manage each request as it moved through a pipeline---something we avoid altogether in \system{}.

Finally, using a dataflow abstraction makes the resulting DAG of computation amenable to a variety of optimizations.
We turn to that topic next.




\section{Optimizing Dataflows} \label{sec:opts}

In this section, we describe the \system{} implementation of each of the optimizations from Section~\ref{sec:bg-opts}. 
All the techniques described in this section are automatic optimizations; the user only needs to select \emph{which} optimizations to enable. 
We return to the topic of automating optimization selection in Section~\ref{sec:conclusion}.
Automatic optimization is a key benefit of the dataflow model, allowing users to focus on pipeline logic, while \system{} takes care of operational concerns including deployment and scheduling.

The \system{} dataflow API produces a DAG of operators selected from Table~\ref{table:api}.
\cloudburst{} provides a lower-level DAG-based DSL for specifying compositions of black-box Python functions~\cite{sreekanti2020cloudburst}. 
A naive 1-to-1 mapping of a user's \system{} DAG into an isomorphic \cloudburst{} DAG will produce correct results,
but we show in this section that we can do much better.

Our compilation occurs at two levels. 
\textbf{Dataflow rewrites} captures static \system{}-to-\system{} optimizations within the dataflow topology, including operator fusion and competitive execution. 
\textbf{Dataflow-to-FaaS compilation} translates \system{} dataflows to \cloudburst{} DAGs that expose opportunities for dynamic runtime decisions in the FaaS infrastructure. 
These include \emph{wait-for-any} DAG execution, autoscaling and hardware-aware function placement, locality-aware scheduling and batching.

In many cases \cloudburst{} was not designed to leverage optimizations apparent at the \system{} layer. 
As a result we had to extend \cloudburst{} in terms both of its API and its runtime mechanisms; we highlight these changes below.

\smallitem{Operator Fusion} \label{sec:opts-fusion}
In operator fusion, a chain of \foperator{}s within a \system{} DAG is encapsulated into a single \fuse{} operator.
The benefit of \fuse{} is that it is compiled into a single \cloudburst{} function; as a result, the \cloudburst{} DAG scheduler places all logic within \fuse{} at a single physical location for execution.
With fusion enabled, \system{} will greedily fuse together all operators in a chain, but will optionally avoid fusing operators with different resource requirements (CPUs vs GPUs).


\smallitem{Competitive Execution} \label{sec:opts-compete}
Competitive execution is used to reduce the tail latency (usually 95th or 99th percentile) of operators with highly variable execution times.
Executing multiple replicas of such operators in parallel significantly reduces perceived latency~\cite{competetiveexecution, paritymodels, sureshc3}.
This is a straightforward dataflow rewrite in \system{}: we create redundant parallel replicas of the operator in question and add an \anyof{} to consume the results. Then the runtime can choose the replica that completes first.

\cloudburst{}'s DAG execution API was ill-suited to \anyof{}: It waits for every upstream function in the DAG before executing a function (\emph{wait-for-all} semantics).
We modified \cloudburst{} to enable functions to execute in \emph{wait-for-any} mode. 



\smallitem{Operator Autoscaling and Placement} \label{sec:opts-scaling}
Each operator in a dataflow may have different performance needs---memory consumption, compute requirements and so on. 
This heterogeneity is particularly pronounced in AI prediction serving. 
For example, an image processing flow might use a CPU-intensive preprocessing stage before a GPU-based neural net.

We seek two optimizations: per-operator autoscaling decisions, and the scheduling of each operator to run on well-chosen hardware.
Continuing our example, if the preprocessing stage was serialized and slow, while the neural net stage was efficient and supported batching, it makes sense to deploy preprocessing functions onto many CPU nodes, and the neural net onto fewer GPU nodes.

\system{}'s dataflow model makes it easy to autoscale dataflow pipelines in a fine-grained way.
As described above, the \system{} compiler translates each operator into a separate \cloudburst{} function (unless fused).
Like other FaaS systems, \cloudburst{} natively autoscales function individually: it adds and removes replicas as load (across many execution requests) changes.

However, \cloudburst{}---like most FaaS systems---does not support specialized hardware today.
We extended \cloudburst{}'s runtime to add support for GPUs in addition to regular CPU-based executors.
Then we extended the \cloudburst{} API to allow functions to be annotated with different resource class labels, and enhanced the \cloudburst{} scheduler to partition its task pool by label. The default \cloudburst{} scheduling policy is applied within each class. It is interesting to consider new policies in this context, but beyond the scope of this paper.




\smallitem{Data Locality via Dynamic Dispatch} \label{sec:opts-locality}
Data retrieval is an important part of many prediction pipelines. 
For example, a recommender system might look at a user's recent click history, then query a database for a set of candidate products before returning a set of recommended items.
Ideally, we want to avoid fetching data over the network, as this can quickly become a latency bottleneck.
\cloudburst{}, by default, performs locality-aware scheduling---it attempts to schedule computation on machines where the task's data dependencies are likely to be cached.
However, its API requires that all of a request's data dependencies be \emph{pre-declared}.
This conflicts with the dynamic nature of many prediction pipelines---in the example above, the candidate product set is determined by the user's recent actions, so the \cloudburst{} scheduler will not know the request's dependencies \emph{a priori}. 
Avoiding this pitfall requires both \system{} rewrites and a \cloudburst{} modification.


We implement two \system{} rewrites.
First, \system{} fuses each \lookup{} operator with the operator downstream from it, so that the processing is colocated with the \lookup{}. 
Second, \system{} rewrites \lookup{} operators to enable \cloudburst{} to perform \emph{dynamic dispatch} of the (fused) operator at a machine that has cached the column value. 
To support this, the column argument to \lookup{} is converted to a reference $ref$ that \cloudburst{} can process at runtime.
The \system{} compiler then splits the dataflow just before the \lookup{}, and generates \emph{two} \cloudburst{} DAGs, flagging the first DAG with a $\mbox{\em to-be-continued}(d, ref)$ annotation, where $d$ is a pointer to the second DAG. 


We then modified the \cloudburst{} runtime to support \emph{to-be-continued} annotations.
When a \emph{to-be-continued} DAG finishes executing in \cloudburst{}, the result---rather than being returned to the user---is sent back to the 
\cloudburst{} scheduler along with a resolved $ref$ (a KVS key) and the DAG ID $d$. 
Given the $ref$, the scheduler can place the DAG $d$ on a machine where the KVS entry is likely to already be cached.




\smallitem{Batching} \label{sec:opts-batching} 
Batching is a well-known optimization for prediction serving, taking advantage of the parallelism afforded by hardware accelerators like GPUs
Many popular ML libraries like PyTorch offer APIs that perform a batch of predictions at once. 
Because our dataflow model is based on small request \ftable{}s, we augment our runtime to form batches across invocations of \texttt{execute}.
To support this, \system{}'s API provides a flag for the function arguments to \map{} and \filter{} to declare batch-awareness.
We then modified the \cloudburst{} API to recognize that flag as we pass it down in compilation. 
Finally we modified the \cloudburst{} executor, so that when it is running a batch-enabled function, it dequeues multiple execution requests 
and executes the entire batch in a single invocation of the function.
The maximum batch size is configurable (defaults to 10). 
The executor demultiplexes the results, and the API returns results for each invocation separately. 




\smallitem{Tradeoffs and Limitations} \label{sec:opts-tradeoffs}
It is worth noting that the optimizations we discuss here have clear tradeoffs.
For example, operator fusion is at odds with fine-grained autoscaling: If a slow function and a fast function are fused, an autoscaler cannot selectively allocate more resources to each function. 
However, automated strategies for trading off these optimizations, such as a cost-based optimizer or a planner, are out of scope for this paper.
We focus on the architecture and mechanisms that enable optimized dataflows; we return to the idea of automated optimization in Section~\ref{sec:conclusion}.
\section{Evaluation} \label{sec:eval}

\begin{figure*}
  \centering
    \includegraphics[width=.95\textwidth]{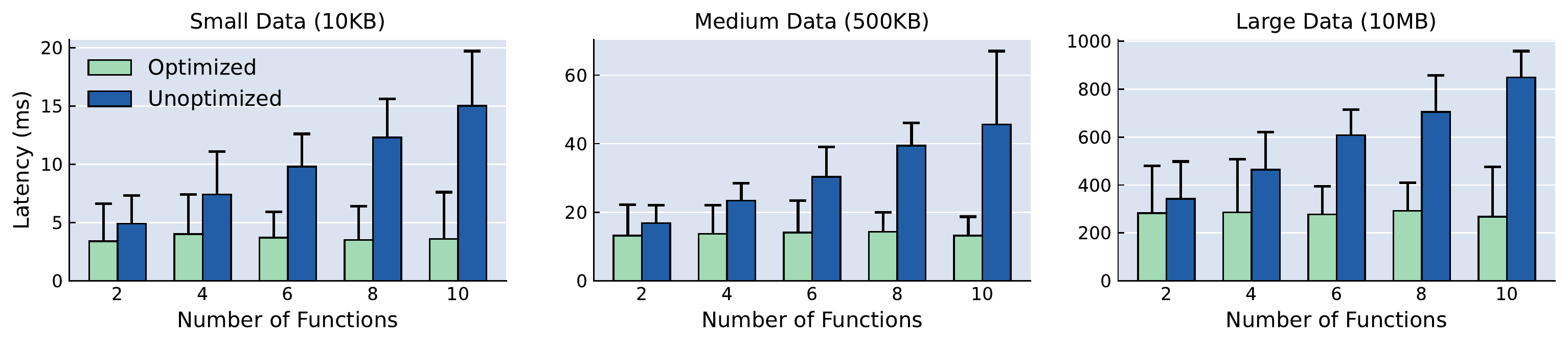}
    \vspace{-1.5em}
  \caption{\small
    Median (bar) and 99th percentile (whisker) latencies for functions chains (length 2 to 10) with varying data sizes (10KB to 10MB).
  }
  \label{fig:fusion}
\end{figure*}

\begin{figure*} 
  \centering
    \includegraphics[width=.95\textwidth]{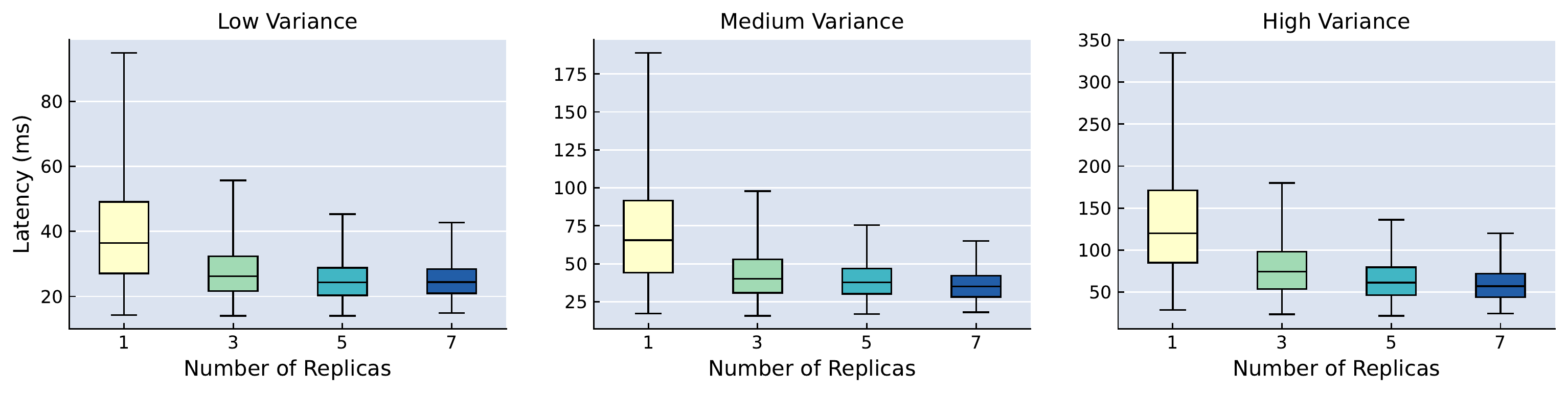}
    \vspace{-1.5em}
  \caption{\small
    Latencies (1st, 25th, 50th, 75th, and 99th percentile) as a function of the number of additional replicas computed of a high-variance function.
  }
  \label{fig:compete}
\end{figure*}

In this section, we present a detailed evaluation of \system{}.
Section~\ref{sec:eval-micro} studies each of the optimizations from Section~\ref{sec:opts} in isolation using synthetic workloads.
We then compare \system{} against state-of-the-art industry and research systems on real-world pipelines. 
We ran all experiments on AWS in the \texttt{us-east-1a} availability zone.
\cloudburst{} CPU nodes used \texttt{c5.2xlarge} instances (2 executors per machine); GPU nodes used \texttt{g4dn.xlarge} instances with Tesla T4 GPUs.


\subsection{Optimization Microbenchmarks} \label{sec:eval-micro}

We evaluate each of the proposed optimizations in isolation on synthetic workloads.  

\subsubsection{Operator Fusion} \label{sec:eval-micro-fusion}

We first study the benefits of operator fusion on linear chains of functions.
The main benefit of fusion is avoiding the cost of data movement between compute locations.
Correspondingly, this experiment varies two parameters: the length of the function chain and the size of data passed between functions.
The functions themselves do no computation---they take an input of the given size and return it as an output.
The output is passed downstream to the next function in the chain.

For each combination of chain length and size, we measure an optimized (fused) pipeline and an unoptimized pipeline.
The fused pipelines execute all operators in a single \cloudburst{} function, while the unfused pipelines execute every stage in a separate \cloudburst{} function.
Figure~\ref{fig:fusion} reports median (bar) and 99th percentile (whisker) latencies for each setting.

As expected, for each input data size, the median latency of the optimized pipelines is roughly constant.
The 99th percentile latencies have slightly more variation---generally expected of tail latencies---but there is no discernible trend. 
The median latencies of the unoptimized pipelines increase linearly with the length of the function chain---primarily because the cost of data movement increases with the length of the chain.
For smaller chains, we see that fusion only imporoves latencies by 20-40\%, depending on data size; however, fusing longer chains  leads to improvements up to 4$\times$.

\textit{\textbf{Takeaway}: Operator fusion in} \system{} \textit{yields up to a 4$\times$ latency decrease by avoiding the overheads of data serialization and data movement between function executors.}

\subsubsection{Competitive Execution} \label{sec:eval-micro-compete}


We proposed competitive execution to reduce latencies for operators with variable runtimes. 
As discussed in Section~\ref{sec:opts-compete}, \system{} executes multiple replicas of a high-variance operator in parallel and selects the result of the replica that finishes first.
Here, we construct a 3-stage pipeline; the first and third operators do no computation.
The second function draws a sample from one of three Gamma distributions and sleeps for the amount of time returned by the sample.
For each Gamma distribution the shape parameter is $k=3$ and the scale parameter is $\theta \in \{1,2,4\}$ corresponding to low, medium, and high variances.

We expect that increasing the number of replicas, particularly for high variance distributions, will noticeably reduce tail latencies.
Figure~\ref{fig:compete} shows our results---the boxes show the interquartile range, and the whiskers show the 1st and 99th percentiles.
In all cases, increasing from 1 to 3 replicas reduces \emph{all} latencies significantly: tail latencies decreased 71\%, 94\%, and 86\% for low, medium, and high variances, and  median latencies decreased 39\%, 63\%, and 62\%.

Beyond 3 replicas, higher variance models see larger improvements.
For the low variance dataflow, increasing from 3 to 7 replicas yields a 30\% improvement in tail latency and only a 7\% improvement at median.
However, for the high variance setting, we observe a 50\% improvement at the tail and a 31\% improvement in median latency.

\textit{\textbf{Takeaway}: Increasing the number of replicas reduces both tail and median latencies, with particularly significant improvements for extremely highly-variable dataflows.}

\subsubsection{Operator Autoscaling} \label{sec:eval-micro-scaling}

\begin{figure}[t] 
  \centering
    \includegraphics[width=\figwidth]{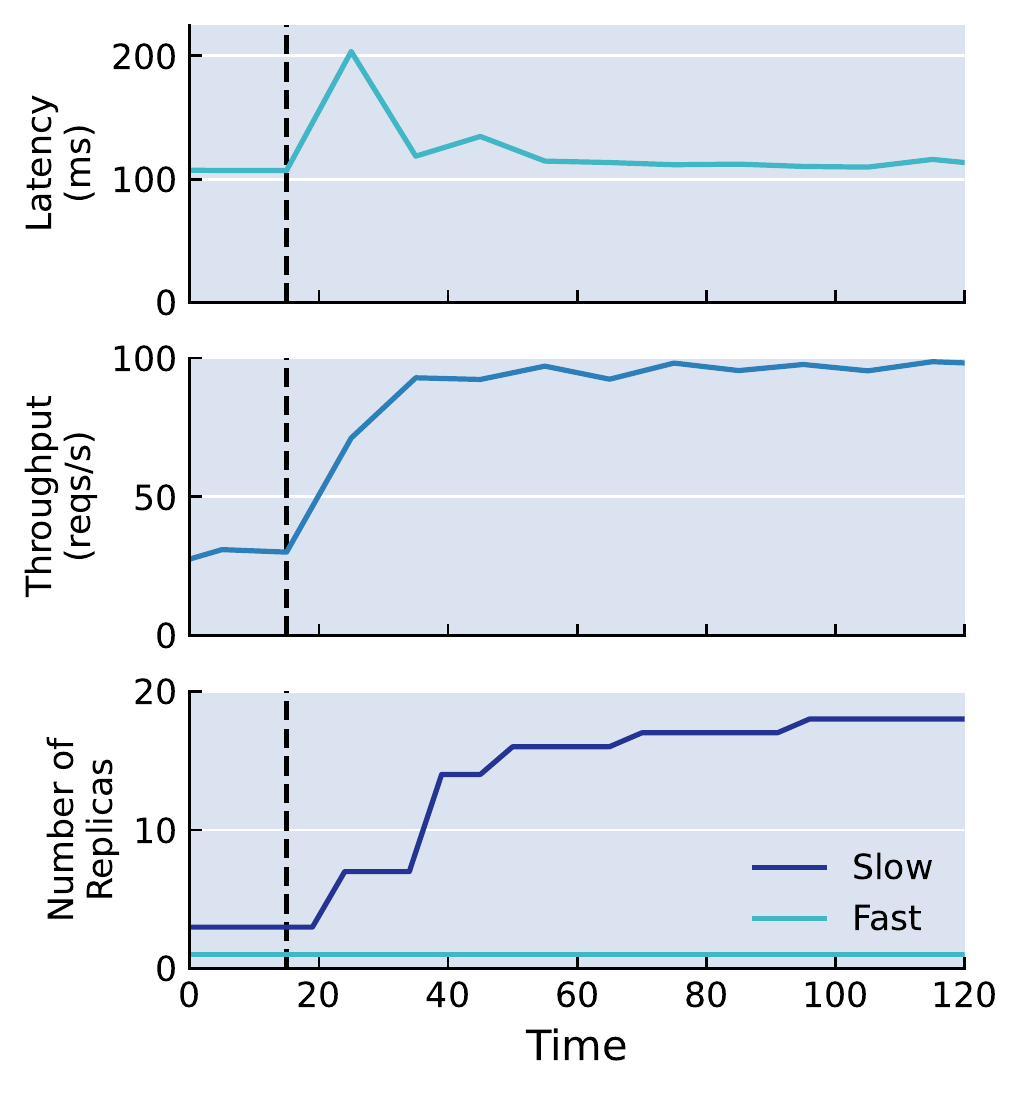}
    \vspace{-3em}
  \caption{\small
  Median latency, throughput, and resource allocation in response to a load spike for a pipeline with a fast and a slow function.
  }
  \label{fig:scaling}
\end{figure}

In this section, we look at \system{}'s and \cloudburst{}'s ability to respond to load changes with fine-grained operator scaling.
We study a workload with two functions---one fast and one slow---similar to the example from Section~\ref{sec:opts}.
We introduce a sudden load spike and measure latency, throughput, and resource allocation.
Figure~\ref{fig:scaling} shows our results.
We begin with 4 client threads issuing requests simultaneously; latency and throughput are steady.
There are 3 threads allocated to the slow function and 1 thread allocated to the fast function.

At 15 seconds, we introduce a 4$\times$ load spike.
Latency immediately increases as the allocated resources are saturated.
The autoscaler responds by adding 16 replicas of the slow function over 15 seconds, and after time 40, latency returns to pre-spike levels (with a minor blip); throughput stabilizes at a new high.
In the meantime, there is no change in the resource allocation of the fast function---it remains at 1 replica.

Over the remaining minute, the autoscaler adds 2 more replicas of the slow function to introduce slack into the system.
This is because the existing resource allocation matches the incoming request rate exactly, and the autoscaler creates a small amount of excess capacity to account for potential future load spikes.
There is no increase in throughput because the existing resources were sufficient to service the incoming request rate. 
This type of fine-grained resource allocation is especially useful for pipelines with heterogeneous resource requirements---a user would not want to scale up GPU allocation beyond necessary for a pipeline is bottlenecked on a CPU task.

\textit{\textbf{Takeaway}:} \system{}\textit{'s dataflow model allows for fine-grained resource allocation in the underlying serverless runtime, enabling nimble responses to load changes while ensuring efficient use of resources.}

\subsubsection{Locality} \label{sec:eval-micro-locality}

\begin{figure} 
  \centering
    \includegraphics[width=\figwidth]{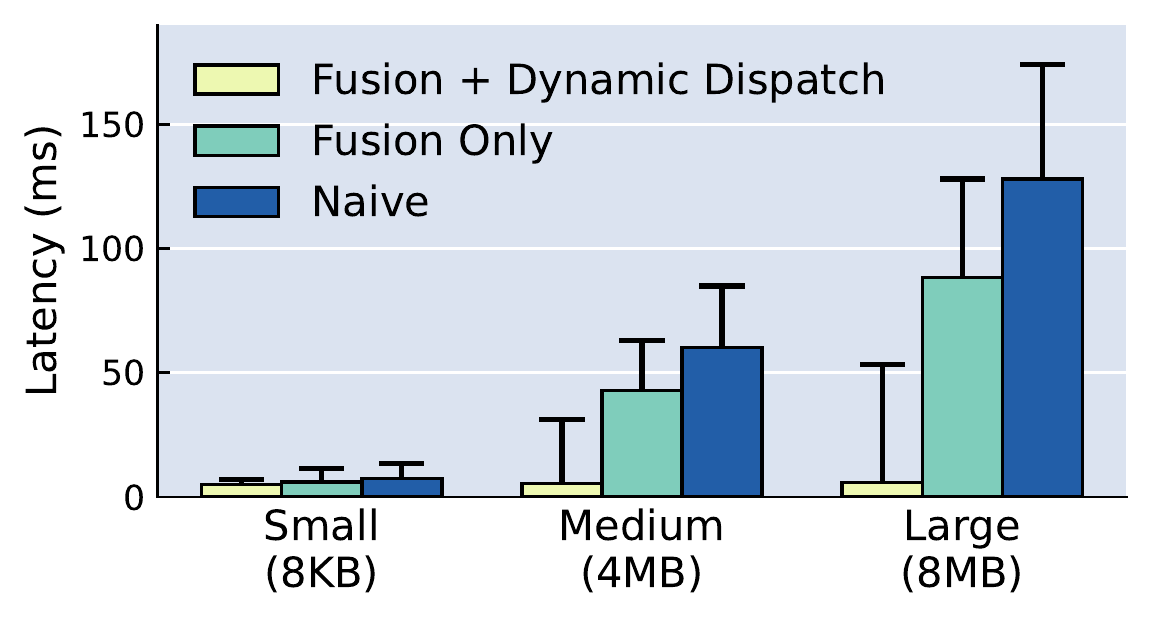}
    \vspace{-2em}
  \caption{\small
  Median latency and 99th percentile latencies for a data-intensive pipeline on \system{} with the fusion and dynamic dispatch optimizations enabled, only fusion enabled, and neither enabled.
  }
  \label{fig:locality}
\end{figure}

Next, we look at the benefits of data locality.
We picked a representative task in which we access each of a small set of objects (100) a few times (10) in a random order.
The pipeline consists of a \map{} that pick which object to access, a \lookup{} of the object, and a second \map{} that computes a result (the sum of elements in an array).
We vary the size of the retrieved data from 8KB to 8MB. 
Figure~\ref{fig:locality} shows our results.
In all settings, we warm up the \cloudburst{}'s caches by issuing a request for each data item once before starting the benchmark.

As described in Section~\ref{sec:opts}, \system{}'s locality optimization has two components: (1) fusing \lookup{}s with downstream operators and (2) enabling dynamic dispatch to take advantage of \cloudburst{}'s scheduling heuristics.
We measure the incremental benefits of both these rewrites.
The Naive bar Figure~\ref{fig:locality} implements neither optimization---data is retrieved from the KVS in the execution of the \lookup{} operator and shipped downstream to the next operator.
The Fusion Only bar merges the \lookup{} with the second \map{} but does not use dynamic dispatch.
The Fusion + Dispatch bar uses both optimizations.

For small data sizes, our optimizations make little difference---the cost of shipping 8KB of data is low, and the Naive implementation is only 2.5ms slower than having both optimizations implemented.
As we increase data size, however, Naive performance significantly worsens---for each request, data is moved once from the KVS to the \lookup{} operator and again from the \lookup{} to the second \map{}.
The Fusion Only operator avoids \emph{one} hop of data movement (between operators), but relies on random chance to run on a machine where its input data is cached---this does not happen very often.
With the dynamic dispatch optimization implemented, \system{} takes advantage of locality-aware scheduling and for the largest data size (8MB) is 15$\times$ faster than Fusion Only and 22$\times$ faster than Naive.
Tail latencies, however, do increase noticeably with data size, as the requests that incur cache misses will still pay data shipping costs.

\textit{\textbf{Takeaway}:} \system{} \textit{'s locality optimizations enable it to avoid data shipping costs and lead to an order of magnitude improvement in latencies for non-trivial data accesses.}


\begin{figure} 
  \centering
    \includegraphics[width=\figwidth]{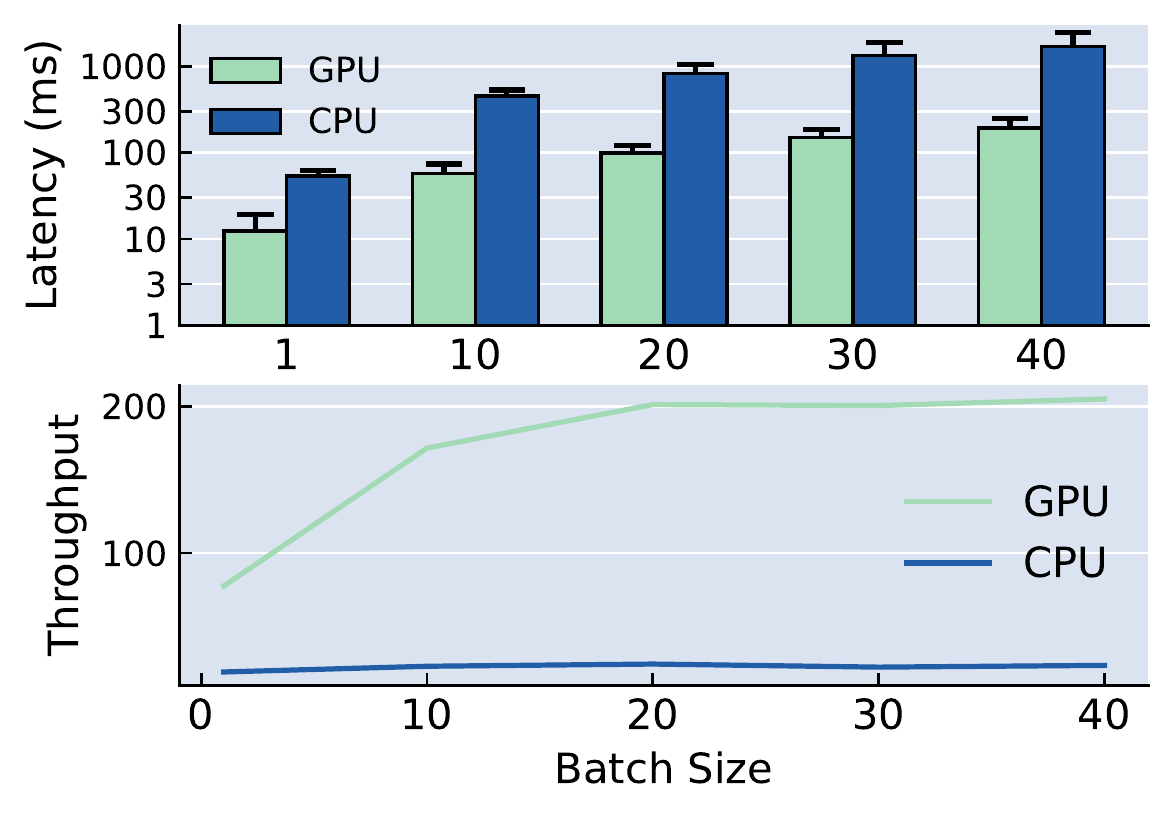}
    \vspace{-2em}
  \caption{\small
    A comparison of CPUs and GPUs on \system{}, measuring latency and throughput while varying the batch size for the ResNet-101 computer vision model.
  }
  \label{fig:batching}
\end{figure}
\subsubsection{Batching} \label{sec:eval-micro-batching}
Finally, we look at the benefits of batching.
In this experiment, we introduce GPUs, since batching primarily benefits with highly parallel hardware.
We execute a pipeline with a single model (the ResNet-50~\cite{he2016deep} image classification model in PyTorch) and no other operators. 
Enabling batching required changing only two lines of user code---using \texttt{torch.stack} to combine inputs into a single tensor.
Our experiment varies the batch size from 1 to 40 in increments of 10.
We asynchronously issue $k$ requests (where $k$ is the batch size) from a single client in order to control the batch size and measure the time until all results are returned.
Figure~\ref{fig:batching} reports latency (on a log-scale) and throughput for each batch size.

At baseline (batch size 1), the GPU has a roughly 4$\times$ better latency and throughput than the CPU.
Increasing the batch size for the CPU from 1 to 10 yields a 20\% increase in throughput (18 to 22 requests) with an 8$\times$ increase in latency.
Beyond that point, latency increases linearly and throughput plateaus---standard CPUs do not support parallel execution.

On a GPU, we see a 4.5$\times$ jump in latency and 2.2$\times$ increase in throughput from size 1 to 10.
Between 10 and 20, we see a further 70\% increase in latency with only an 18\% increase in throughput. 
Past size 20, the GPU is saturated; throughput plateaus and latency increases linearly with batch size.
Note that between batch sizes 1 and 20, there is only an 8$\times$ increase in latency (\~12ms to ~100ms)---well within the bounds for an interactive web application~\cite{he2012zeta}---while significantly improving the throughput (3$\times$).

\textit{\textbf{Takeaway}: \system{}'s batching optimization enables a 3$\times$ increase in throughput while retaining interactive latency.}

\subsection{Prediction Serving Pipelines} \label{sec:eval-pipelines}

Having measured \system{} in isolation, we now look at real prediction serving pipelines and compare \system{} to state-of-the-art commercial and research systems: AWS Sagemaker and Clipper~\cite{crankshaw2017clipper}.
AWS Sagemaker is a hosted, end-to-end platform for managing machine learning models that includes a containerized deployment tool.
We selected it as a representative of a number of hosted model management tools with prediction serving capabilities.
Clipper is an open-source research platform that automatically containerizes and deploys models while maintaining strict latency goals.
We measure \system{} against both systems on four real world pipelines: an image cascade, a video processing pipeline, a neural machine translation task, and a recommender system.

For each pipeline, we sampled multiple optimization strategies on \system{}.
Due to space limits, we do not describe results from each configuration; we limit ourselves to the optimization strategies that worked best and the corresponding results. 
We return to automating the optimzation process in Section~\ref{sec:conclusion}.

\subsubsection{Pipelines} \label{sec:eval-pipelines-defs}


\begin{figure}[b]
\begin{python}
 flow = cloudflow.Dataflow([('img', numpy.ndarray)])
 img = flow.map(preprocess)
 resnet = img.map(resnet_101)
 inception = resnet.filter(low_confidence).map(inception_v3)
 label = resnet.join(inception).map(pick_best_prediction)
 
 category_counts = people.union(vehicles).groupby('category').agg('count')
 label = category_counts

 flow.output = label
\end{python}
\caption{The \system{} implementation of the image cascade pipeline in Section~\ref{sec:eval-pipelines}.} 
\label{lst:img-cascade}
\end{figure}

\begin{figure}[b]
\begin{python}
 flow = cloudflow.Dataflow([('frame', numpy.ndarray)])
 yolo_preds = flow.map(yolov3)
 people = yolo_preds.map(resnet101_people)
 vehicle = yolo_preds.map(resnet101_vehicles)
 category_counts = people.union(vehicles).groupby('category').agg('count')
 label = category_counts

 flow.output = label
\end{python}
\caption{The \system{} implementation of the video stream processing pipeline in Section~\ref{sec:eval-pipelines}.} 
\label{lst:yolo}
\end{figure}

\begin{figure}[b]
\begin{python}
 flow = cloudflow.Dataflow([('english', str), ('other', str)])
 language_type = flow.map(yolov3)
 french = language_type.filter(is_french).map(translate_from_french)
 german = language_type.filter(is_german).map(translate_from_german)
 result = german.join(french)
 
 flow.output = result
\end{python}
\caption{The \system{} implementation of the neural machine translation pipeline in Section~\ref{sec:eval-pipelines}.} 
\label{lst:nmt}
\end{figure}

\begin{figure}[b]
\begin{python}
 flow = cloudflow.Dataflow([('uid', int), ('recent_clicks', numpy.ndarray)])
 topk_for_user = flow.lookup('uid', type='column_ref') \
                     .map(pick_category) \ 
                     .lookup('category', type='column_ref')
                     .map(get_topk)
 
 flow.output = topk_for_user
\end{python}
\caption{The \system{} implementation of the recommender system pipeline in Section~\ref{sec:eval-pipelines}.} 
\label{lst:recsys}
\end{figure}

\smallitem{Image Cascade}. A cascade pipeline (described in Section~\ref{sec:api-control-flow}) chains together models of increasing complexity, executing later models only if the earlier models are unsure of their predictions.
We use a cascade with two models: ResNet-101~\cite{he2016deep} and Inception v3~\cite{inception-v3}.
We preprocess an input image and execute ResNet; if ResNet's confidence is below 85\%, we then execute Inception and pick the result with the higher confidence. 
We run this pipeline with a random sample of images from the ImageNet dataset~\cite{russakovsky2015imagenet}.

\smallitem{Video Streams}. Processing video streams is an extremely data- and compute-intensive task.
We use a standard video processing flow \cite{videoanalytics} that uses YOLOv3~\cite{redmon2018yolov3} to classify the frames of a short video clip (1 second).
Based on the YOLO classifications, we select frames that contain people and vehicles (representative of a pipeline for an autonomous vehicle) and send those frames to one of two ResNet-101 models. 
The two ResNet models are executed in parallel and more specifically classify people and vehicles, respectively.
We union the results of the two models, group by the ResNet classifications, and \texttt{count} the occurrences of each class per input clip.

\smallitem{Neural Machine Translation}. Models that translate text between languages are becoming increasingly widely used (e.g., chat applications, social networks).
These models are also notoriously large and computationally intensive.
We model a translation task by first classifying an input text's language using fastText, an open-source language modeling library~\cite{joulin2016bag, joulin2016fasttext}.
We limit our inputs to French and German.
Based on the language classification, we then feed a second input into one of two models that translates between English and the classified language. 
The translation models are implemented in the PyTorch FAIRSEQ library~\cite{fairseq-library} based on sequence learning translation models~\cite{language-models}.

\smallitem{Recommender System}. Recommender systems are ubiquitous among internet services.
Facebook recently published a case study of their DNN-based personalized recommenders~\cite{gupta2020architectural}; we implement a pipeline modeled after their architecture. 
A request comes in with a user ID and recent items the user has clicked on; based on the set of recently clicked items, we generate a product category recommendation.
The pipeline then computes a top-$k$ set of items from that product category, using the user's pre-calculated weight vector.
This is done with a simple matrix multiplication-based scoring scheme.
While the user vectors are small (length 512), the product category sets can be large (roughly 10MB)---as a result, enabling locality and minimizing data movement is an important part of implementing these pipelines efficiently.


\subsubsection{Benchmark Setup} \label{sec:eval-pipelines-setup}

\smallitem{\system{}}. For each pipeline, we start with one replica per operator and run a 200 request warm-up phase that allows the \cloudburst{} autoscaler to settle on a resource allocation.
We then run 1,000 requests from 10 benchmark clients in parallel and report median latency, 99th percentile latency, and throughput below.
Unless stated otherwise, we copied the exact resource allocation from \system{} to each of the other systems.

\smallitem{Sagemaker}. Sagemaker requires users to provide a Docker container per pipeline stage---a cumbersome and time-consuming process.
We wrap each pipeline stage in a bespoke container and deploy it as a Sagemaker endpoint.
We do not rely on Sagemaker's built-in inference pipelines feature, as it does not support parallelism, which is required in our workloads.
Instead, we built a proxy service that manages each client request as it passes through the pipeline.
This enables us to invoke multiple endpoints in parallel when necessary.
CPU workers in Sagemaker used \texttt{ml.c5.2xlarge} instances, and GPU workers used \texttt{ml.g4dn.xlarge} instances.


\smallitem{Clipper}. 
Clipper is a state-of-the-art research system that, similar to Sagemaker, deploys models as microservices; Clipper does not support pipelines in any form. 
Similar to our Sagemaker deployments, we use custom code to move each request through the pipeline. 
Deploying models using Clipper's API was easier than with Sagemaker, but certain pipelines with bespoke dependencies required us to build custom Docker containers.
The simplest way to deploy model replicas on Clipper was to create multiple replicas of the model on a large machine---however, we ensured that the number of resources (vCPUs, RAM, and GPUs) per worker was the same as on \system{} and Sagemaker.
Note that Clipper also supports batching, which we enabled for GPU workloads.

\subsubsection{Results} \label{sec:eval-pipelines-results}

\begin{figure*} 
  \centering
    \includegraphics[width=.95\textwidth]{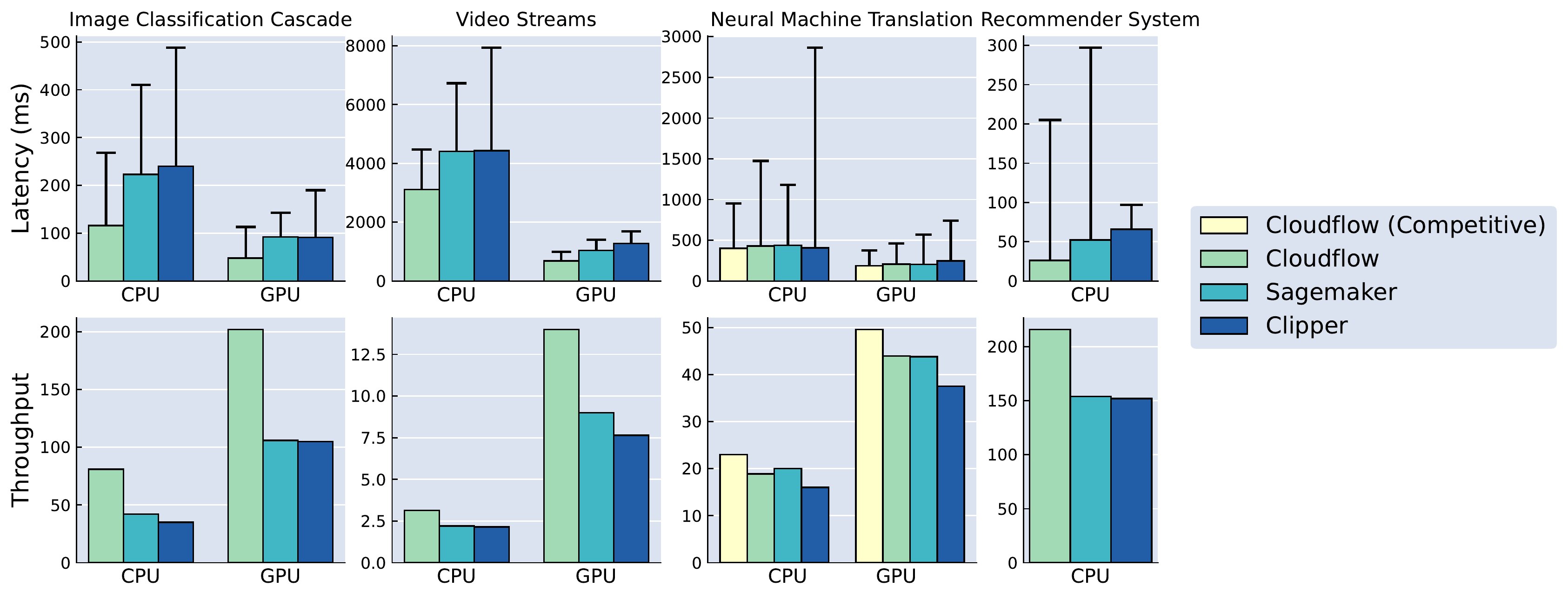} 
    \vspace{-1.5em}
  \caption{\small
    Latencies and throughputs for each of the four pipelines described in Section~\ref{sec:eval-pipelines-defs} on \system{}, AWS Sagemaker, and Clipper.
  }
  \label{fig:e2e}
\end{figure*}

We report results for each pipeline in Section~\ref{sec:eval-pipelines-defs} on both CPU and GPU deployments, except for the recommender system, which we only run on CPU.
We omit GPU results for this pipeline because, as noted by \cite{gupta2020architectural},  recommendation models have low arithmetic intensity and do not usually benefit from GPU acceleration.
Unless stated otherwise, we enabled batching on \system{} and Clipper for all GPU workloads and disabled it for all CPU workloads, as per our results in Section~\ref{sec:eval-micro-batching}.

\smallitem{Image Cascade}.
The leftmost graph in Figure~\ref{fig:e2e} shows results for the image cascade.
We enabled \system{}'s fusion optimization and merged the whole pipeline into a single operator.
Despite having a mix of CPU and GPU operators in the pipeline, we found that the CPU execution costs were low (10-15ms) and that reducing data movement best improved \system{}'s performance.

On both CPU and GPU deployments, \system{} has a roughly 2$\times$ better median latency than the other systems.
This translates into a 2$\times$ increase in throughput for both hardware configurations.
These improvements are largely due to reduced data movement costs enabled by operator fusion.
We also note the benefits of batching by comparing Clipper and Sagemaker's performance.
In CPU mode where neither system uses batching, Sagemaker outperforms Clipper by 20\%; however, in GPU mode, where Clipper supports batching and Sagemaker does not, Clipper is able to close the performance gap and \emph{match} Sagemaker's throughput.



\smallitem{Video Streams}. The video streaming pipeline was the most data intensive: Each video input had 30 frames, which was about 20MB of data.
In this pipeline, there were only GPU models (YOLOv3 and ResNet-101) without any CPU stages---we again apply operator fusion to the whole pipeline to generate a single \cloudburst{} function.
We found that the cost of executing the two parallel ResNet instances in serial was lower than the cost of shipping the input over the network.

CPU deployments are expectedly slow---\system{}'s median latency was 3.1 seconds, while Sagemaker's and Clipper's were around 4.4 seconds; tail latencies were 4.4, 6.7, and 7.9 seconds respectively.
\system{}'s reduced data movement enables significant improvements (41\% at median and up to 80\% at the 99th percentile) but was still capped at 3.1 requests per second.

Results are much more promising for GPU deployments.
\system{}'s median latency was 685ms, and its 99th percentile latency was about 1 second (996ms)---meaning that \system{} can consistency process video in real-time.
Sagemaker and Clipper were both slower, with median latencies of 1 and 1.3 seconds, respectively, and 99th percentile latencies at 1.4 and 1.7s seconds.
\system{} is able to process 14 requests per second,  while the other systems were both below 9 requests a second.

\smallitem{Neural Machine Translation}. The neural machine translation task was the best case scenario for both Clipper and Sagemaker as it had the least input data (strings to translate) while being very computationally intensive.
We found that the NMT models---particularly on CPU---had high variance in runtimes, so we enabled competitive execution in \system{}.
However, since competitive execution required allocating extra resources not available to Sagemaker and Clipper, we report \system{} measurements with and without competitive execution.


On CPUs, the median latencies for all configurations are similar---Clipper was about 7\% faster than Sagemaker and \system{} without competition.
At the tail, Clipper was the slowest with a 99th percentile of 2.8 seconds; \system{}'s tail latency was 1.5 seconds, and Sagemaker's was 1.2 seconds.
Despite a 25\% lower tail latency, Sagemaker's throughput was only 5\% higher than \system{}'s.
Adding two competitive replicas reduced \system{}'s median latency by 7\% and its 99th percentile by over 50\%. 
This led to a 20\% improvement in throughput compared to \system{} without competition and 15\% improvement over Sagemaker.


On the GPU deployment, \system{} without competition and Sagemaker had similar performance. 
Clipper's median latency was about 20\% higher, likely because its scheduler constructs batches more aggressively than \system{} does.
\system{} without competition and Sagemaker had the same throughput, while Clipper's was 15\% lower.
Introducing competitive GPU replicas lowered median latencies by 10\% and lowered tail latency by almost 25\% relative to \system{} without competition (and by over 50\% relative to Sagemaker).
This translated into a 13\% improvement in throughput.


\smallitem{Recommender System}.
Finally, we look at the recommender pipeline.
This workload was different from the others because it was data-intensive and computationally light---each request required only a matrix multiplication followed by a few simple operations.
The user weight vectors and product categories in this pipeline are pre-generated.
That provides an opportunity for us to exercise \system{}'s locality aware scheduling.

Running on \cloudburst{} allows us to automatically access data in Anna; however, Sagemaker blocks inbound network connections, which Anna requires, as it is an asynchronous KVS.
As a workaround we used AWS ElastiCache's hosted Redis offering, which has similar performance to Anna \cite{wu2019anna}, as the storage engine for Clipper and Sagemaker.
To simulate the caches in \cloudburst{}, we also added in-memory caches of size 2GB to Sagemaker and Clipper, that allowed them to cache user weight vectors and product categories---however, these systems do not have \system{}'s dynamic dispatch, so the likelihood of a cache miss is much higher.
We pre-generate 100,000 user weight vectors (length 512, 4KB each) and 1,000 product categories (2,500 products per category, 10MB each) and put them in storage.
For each request, we input a random user ID and a random set of recent clicks---this translates to a random product category per request.



\system{}'s locality optimizations lead to a 2$\times$ improvement in median latency over Sagemaker's and 2.5$\times$ improvement over Clipper's.
\system{} reliably places requests on machines with cached inputs, whereas Sagemaker and Clipper are much less likely to see cache hits.
Tail latencies for all systems are high due to the cost of retrieving data over the network
Nonetheless, the lower median latencies translate to a roughly 40\% increase in throughput over Sagemaker and Clipper.

\textit{\textbf{Takeaway}:\system{}'s optimizations enable it to outperform state-of-the-art prediction serving systems up to 2$\times$ on real-world tasks; importantly, it is able to meet real-time latency goals for a data- and compute-intensive task like video streaming.}
\section{Related Work} \label{sec:related}

\smallitem{Dataflow Abstractions}.
Dataflow abstractions have deep roots in both software and hardware architectures (e.g., ~\cite{dennis1974first,dennis1974preliminary}). In the last decade, dataflow emerged as a core programming abstraction for systems such as Spark~\cite{10.1145/2934664} and Dryad~\cite{10.1145/1272996.1273005} that excel in high throughput big data processing.
Other systems like Naiad~\cite{10.1145/2517349.2522738}, Flink~\cite{carbone2015apache}, and Noria~\cite{222635, schwarzkopf} implement stateful dataflow that targets scenarios where the input data continuously streams into the system.
\system{}, in comparison, is designed for handling interactive model serving queries at low latency, and the input of each query is bounded.
In addition to standard dataflow optimizations such as operator fusion, \system{} implements a variety of custom optimizations that are well-suited to interactive model serving, such as competitive execution and fine-grained autoscaling.

\smallitem{Packet Processing and Distributed Systems.} 
Programmable packet processors for networks share our latency focus, but target much less compute-intensive tasks than model serving, using lower-level optimizations. The Click router~\cite{kohler2000click} is an influential example; subsequent work in declarative networking~\cite{loo5} and packet processing languages~\cite{bosshart2014p4} provide higher-level abstractions that often
compile down to dataflow. Similar languages have targeted distributed systems~\cite{alvaro2011consistency,meiklejohn2015lasp,milano2019tour}, with a focus on orthogonal concerns like coordination avoidance.




\smallitem{Lightweight Event Flows.} Systems such as SEDA \cite{welsh2001seda} adapt the resource allocation of different stages of a pipeline based on demand, but do not have a semantic notion of the tasks' meanings and thus cannot perform logical rewrites of the pipeline. 
Additionally, SEDA is focused on single-server rather than distributed executions and has no autoscaling.

\smallitem{ML Optimizations.} TASO~\cite{taso} and ParM~\cite{paritymodels} introduce optimizations on individual models that can be integrated into prediction serving systems. 
TASO performs low-level rerwites of DNN computation graphs to improve model performance.
ParM reduces the tail latencies and enables failure recovery for variable models by using erasure coding to reconstruct slow and failed predictions.
Neither system offers end-to-end optimizations similar to \system{}'s.

\smallitem{Prediction Serving Systems}. Clipper~\cite{crankshaw2017clipper}, TFServe~\cite{tfx}, Azure ML~\cite{pmlr-v50-azureml15}, and AWS Sagemaker either have no support for pipelines or impose restrictions on the topology of the pipelines.
As such, none of these systems perform pipeline optimizations as in \system{}.
Other systems such as Inferline~\cite{inferline} and Pretzel~\cite{lee2018pretzel} are complementary to \system{}; Inferline focuses on efficient scheduling to meet latency goals, and Pretzel exploits white-box pipelines to optimize individual stages. 
We plan to explore both of these approaches in future work.


\section{Conclusion and Future Work} \label{sec:conclusion}

In this paper, we presented \system{}, a dataflow framework for prediction serving pipelines. 
\system{} offers a familiar dataflow programming API that allows developers to easily construct prediction pipelines in a few lines of code. 
We transparently implement two kinds of optimizations over the resulting dataflow graph: dataflow rewrites and dataflow-to-FaaS compilation. 
\system{} deploys pipelines on top of \cloudburst{}, a low-latency, stateful Function-as-a-Service platform; we extended \cloudburst{} with key features to support our workloads.

Our evaluation shows that \system{}'s automated optimizations enable us to outperform state-of-the-art prediction serving systems by up to 2$\times$ for  real-world tasks like image cascades and video processing.
Our architecture and optimization scheme suggest a number of interesting avenues for future work.

\smallitem{Automated Optimization Selection}. Currently, we require manual selection of optimizations best-suited to each workload.
There is a rich history of automated optimization of programs---including dataflows---in the database, systems and programming languages literature.
Techniques like cost-based database query optimization would seem to be a good fit to our API, but will require research to address our heterogeneous operators, hardware, and latency goals.


\smallitem{Meeting Latency SLAs}. We have focused in this work on reducing latency, but we have not thus far considered techniques for managing Service Level Agreements (SLAs) or Objectives (SLOs) regarding latency deadlines. 
Having explicit deadlines---with cost penalties for missing them---brings multi-objective tradeoffs into the picture, which makes the optimization problems we consider more complex.


\smallitem{Verifying Dataflow Correctness}. As we alluded to in Section~\ref{sec:api-dataflow}, \system{} provides runtime type checking of dataflows.
However, there are other changes in machine learning pipelines---e.g., a video monitoring camera inadvertently is turned to face a wall---that can lead to failures without type errors.
TFX~\cite{tfx} helps detect these issues by capturing statistics about the inputs to its models; we plan to bring such capabilities into the existing type monitoring in \system{}.
We also plan to extend the debuggability of \system{} by adding anomaly detection over the passively captured statistics about inputs and outputs.


\bibliographystyle{abbrv}
\bibliography{references}

\end{document}